\documentclass[prl,twocolumn,superscriptaddress]{revtex4}
\usepackage{amssymb,amsmath,amsthm,graphicx}
\usepackage{latexsym}
\usepackage{bm}
\usepackage[]{hyperref}
\usepackage{cleveref}
\usepackage[utf8]{inputenc}
\usepackage{amssymb}
\usepackage{xcolor}

\allowdisplaybreaks
\pdfoutput=1

\begin{document}
\title{Holographic Signatures of Critical Collapse}

\author{Paul~M.~Chesler}
\email{pchesler@g.harvard.edu}
\affiliation{
Black Hole Initiative, Harvard University, Cambridge, MA 02138, USA}
\author{Benson~Way}
\email{benson@phas.ubc.ca}
\affiliation{Department of Physics and Astronomy, University of British Columbia, 6224 Agricultural Road, Vancouver, B.C., V6T 1W9, Canada}

\begin{abstract}
Critical phenomena in gravitational collapse exhibit the universal features of self-similarity, critical scaling, and the appearance of a naked singularity.  We study critical collapse in AdS, focusing on holographic field theory observables.  We demonstrate that the echoing period, critical exponent, and signatures of the naked singularity can all be extracted from the holographic stress tensor.
\end{abstract}

\maketitle

{\bf~Introduction --} Nearly 30 years ago, Choptuik \cite{Choptuik:1992jv} discovered that gravitational systems on the threshold of collapse exhibit universal features (see \cite{Gundlach:2007gc} for a review).  First, as initial conditions are tuned towards criticality, the mass of the produced black hole scales like
\begin{equation}
\label{eq:scaling}
M_H \sim (\epsilon - \epsilon_*)^{\gamma},
\end{equation}
where $\epsilon$ parameterizes the initial data, with a black hole forming for supercritical data $\epsilon > \epsilon_*$.  The exponent $\gamma$ is universal in that it is independent of initial conditions. Second, the critical solution itself exhibits self-similarity that can either be continuous or discrete. For discrete self-similarity, for suitably chosen time and radial coordinates $t$ and $r$, the critical solution is periodic in $\ln r$ and $\ln (t - t_c)$ with a universal period $\Delta$. This ``scale echoing," means that as $t \to t_c$, the critical solution oscillates on increasing finer scales with a naked singularity at $t = t_c$.

Holographic duality \cite{Maldacena:1997re,Gubser:1998bc,Witten:1998qj,Aharony:1999ti} relates quantum theories of gravity to quantum field theories (QFT) without gravity in lower dimensions. {In the limit of a large number of colors $N$ and large `t Hooft coupling $\lambda$, the dual gravitational theory is classical Einstein gravity.}  The most widely studied example of holographic duality is AdS/CFT, which relates gravitational dynamics in anti-de Sitter (AdS) space to the dynamics of a conformal field theory (CFT) that lives on the AdS boundary. Gravitational collapse has a natural QFT interpretation as thermalization: black holes have $O(N^2)$ entropy \cite{Witten:1998zw},  at least $O(\sim N^2)$ lifetime, and yield dual correlation functions that satisfy the fluctuation dissipation theorem \cite{CaronHuot:2011dr,Chesler:2011ds,Chesler:2012zk}.

The ubiquity of critical collapse together with holographic duality suggests that the dynamics of quantum systems on the threshold of thermalization show universal features.  A natural question arises: within the framework of holographic duality, how do critical collapse and its associated scale echoing and scaling relation \ref{eq:scaling} manifest themselves in QFT observables?  While critical phenomena has been studied in AdS \cite{Pretorius:2000yu,Husain:2002nk,Olivan:2015fmy,Santos-Olivan:2016djn,Bizon:2011gg}, and attempts have been made to compute $\gamma$ from the QFT side \cite{Bland:2007sg,AlvarezGaume:2008qs}, the dynamics of QFT observables themselves remain largely unexplored.

We seek to fill this gap by computing the expectation value of the QFT stress tensor $\langle T^{\mu \nu} \rangle$ from a dual gravity state near criticality.  We focus on dynamics in global AdS$_5$, dual to $\mathcal N = 4$ supersymmetric Yang-Mills theory on $\mathbb{R}\times S^3$.  We find that $\langle T^{\mu \nu} \rangle$ exhibits scale echoing. For both subcritical and supercritical solutions, we find that scale echoing in $\langle T^{\mu \nu} \rangle$ terminates when the echo frequency (in linear boundary time $t$) is $f \sim |\epsilon - \epsilon_*|^{-\gamma/2}$. Moreover, we find evidence that as $\epsilon \to \epsilon_*$, scale echoing terminates with $\langle T^{\mu \nu} \rangle$ diverging, meaning the naked singularity in the gravity description manifests itself as a singular stress tensor in the dual QFT.

{\bf~Gravitational description --}  Most studies of critical collapse in AdS utilize a massless scalar field in spherical symmetry. While we have obtained results within this model, the echoing period is rather large, which limits the number of echoing periods that can be captured by numerical time evolution.  We therefore present results from a different model with a shorter echoing period.  Our results for the massles scalar are qualitatively similar.

The model we have chosen actually lies in a theory that is more universal than the massless scalar: pure gravity. We break the spherical $SO(4)$ symmetry of global AdS$_5$ by a homogeneous squashing of the sphere that preserevs a $SO(3)\times U(1)$ symmetry.  This squashing becomes a dynamical degree of freedom which evades Birkhoff’s theorem.  Previous studies in the asymptotically flat version of this setup \cite{Bizon:2005cp} show that the  Minkowski and Schwarzschild solutions serve as endpoints in the evolution of generic initial data, and that critical phenomena can be found by fine tuning initial data. Like the massless scalar in spherical symmetry, this model also has a universal critical exponent and discrete self similarity.

To obtain the boundary observables of both supercritical and subcritical collapse, we must propagate the signals associated to the formation of a horizon out to the AdS boundary.   It is therefore necessary to use a gauge that allows for evolution past horizon formation.  For this reason, rather than the more common radial Schwarzschild-like gauge, we use a maximal slicing gauge, where the trace of the extrinsic curvature $K=0$.

We now describe our metric ansatz:
\begin{align}
\label{metricansatz}
\mathrm ds^2 &= \textstyle \frac{1}{(1-\rho^2)^2}\bigg\{{-}\alpha^2\left(1-\rho^2(2{-}\rho^2)\frac{\beta^2}{a}\right)\mathrm dt^2+ \frac{4\alpha\beta}{a}\rho\,\mathrm dt\mathrm d\rho \nonumber\\
 & \textstyle + \frac{4 {\mathrm d}\rho^2}{a (2 - \rho^2)}+\frac{\rho^2(2{-}\rho^2)}{b^2}\left  [\left(\mathrm d\psi{+}\cos^2(\tfrac{\theta}{2})\mathrm d\phi\right)^2 +\frac{b^3}{4} d\Omega_2^2\right]\!\! \bigg\}\;,
\end{align}
where we have set the AdS radius to unity. $\psi\in[0,2\pi)$, $\phi\in[0,2\pi)$, and $\theta\in[0,\pi)$ are angular directions and $d\Omega^2_2 = \mathrm d\theta^2{+}\sin^2\theta\,\mathrm d\phi^2$ is the metric on the unit 2-sphere.  The functions $\alpha$, $\beta$, $a$ and $b$, depend only on time $t$ and radial coordinate $\rho$. The origin of the geometry lies at $\rho = 0$ and the boundary lies at $\rho = 1$. When $\alpha=a=b=1$ and $\beta=0$, the metric (\ref{metricansatz}) is that of global AdS$_5$. Correspondingly, at $\rho = 1$ we impose the boundary condition $\alpha=a=b=1$ and $\beta=0$, so the geometry is asymptotically AdS, with a boundary metric
\begin{equation}
\label{bndrymetric}
ds_\partial^2 = - dt^2 + d \Omega_3^2,
\end{equation}
with $d \Omega_3^2$ the metric on the unit 3-sphere. At the origin, we impose regularity.

For initial data, we take
\begin{align}\label{initialdata}
&b\big|_{t=0}=0 , && \textstyle
\frac{\partial_tb}{\alpha\sqrt{a}}\big|_{t=0}=\epsilon \exp\left[-\frac{4\rho^2(2-\rho^2)}{\pi^2\sigma^2(1-\rho^2)^2}\right]\;,
\end{align}
with fixed $\sigma=0.1$ and let $\epsilon$ parameterize our initial data. The remaining metric components are determined by constraint equations. From studies of AdS instability \cite{Bizon:2011gg}, this family of initial data is expected to contain an infinite number of critical $\epsilon_*$, corresponding to the number of times the configuration bounces back and forth between the origin and the boundary before eventually approaching the critical solution.  For simplicity, we take $\epsilon$ to be near the highest critical $\epsilon_*$, so that the first signal we see at infinity corresponds to critical behavior.

The holographic stress tensor $\langle T^{\mu}_{\ \nu} \rangle$ is determined by the near-boundary asymptotics of the metric \cite{deHaro:2000vlm}. The stress tensor is covariantly conserved with respect to the boundary metric (\ref{bndrymetric}), which just reflects energy and momentum conservation in the dual field theory. Since the metric functions $\alpha, \beta, a,b$ do not depend on angles, $\langle T^{\mu}_{\ \nu} \rangle$ only depends on time. It follows that $\langle T^{\mu}_{\ \nu} \rangle$ has a single dynamical degree of freedom which, together with the conserved energy, determines all other components of the stress tensor.  We will take the this degree of freedom to be the normalized pressure anisotropy
\begin{equation}
\textstyle\langle B\rangle\equiv2\pi G_5\left(\langle T^{\psi}_{\ \psi}\rangle-\frac{1}{3}\sum_{i\in{\{\psi,\phi,\theta\}}}\langle T^i{}_i\rangle\right),
\end{equation}
where $G_5 = \frac{\pi}{2N^2}$ is the five dimensional Newton constant. Since $\langle T^{\mu}_{\ \nu} \rangle$ is order $N^2$, $\langle B\rangle$ is independent of $N$ in the large $N$ limit. In terms of the bulk geometry,
\begin{equation}
\label{stress2}
\langle B \rangle  = \lim_{\rho \to 1} {\textstyle \frac{1}{64 (1 - \rho)^3} \frac{\partial b}{\partial \rho}}.
\end{equation}

Now we broadly describe our numerical evolution scheme. We are studying a sub-sector of the model in \cite{Choptuik:2017cyd} with the same setup, nomenclature, and numerical methods.  We therefore refer the reader to \cite{Choptuik:2017cyd} for further details and for numerical validation of the code.

The Einstein equation and maximal slicing gauge condition $K=0$ yield a second-order wave equation for $b$.  There are also first-order spatial constraint equations for $\beta$, $a$ and $\alpha$, as well as temporal constraint equations for $a$ and $\beta$.  There is also one spatial constraint equation and one temporal constraint equation for $b$, obtained by reducing its wave equation to first order.

Our main evolution system consists of the wave equation for $b$ (in first-order form), and the temporal constraint for $b$.  The spatial constraint for $b$ is used as added constraint damping terms to improve numerical stability. Evolution proceeds by a fourth order Runge-Kutta method (RK4).  Spatial constraints, which consists of a sequence of linear problems, are solved via Gaussian elimination at each RK4 step for $\beta$, $a$ and $\alpha$.  The remaining equations are not solved directly and verified afterwards.  For all data presented here, relative energy violation and constraint violation are within or below $10^{-6}$.

Spatial discretization is supplied by a spectral element mesh with Legendre-Gauss-Lobatto nodes, using a discontinuous Galerkin method with Lax-Friedrichs flux to handle inter-element coupling.  We utilise adaptive mesh refinement to maintain numerical control over the large gradients in the fields.  When an apparent horizon forms, a portion of the grid inside the horizon is excised before continuing with the numerical evolution.

\begin{figure}
\centering
\includegraphics[width=.45\textwidth]{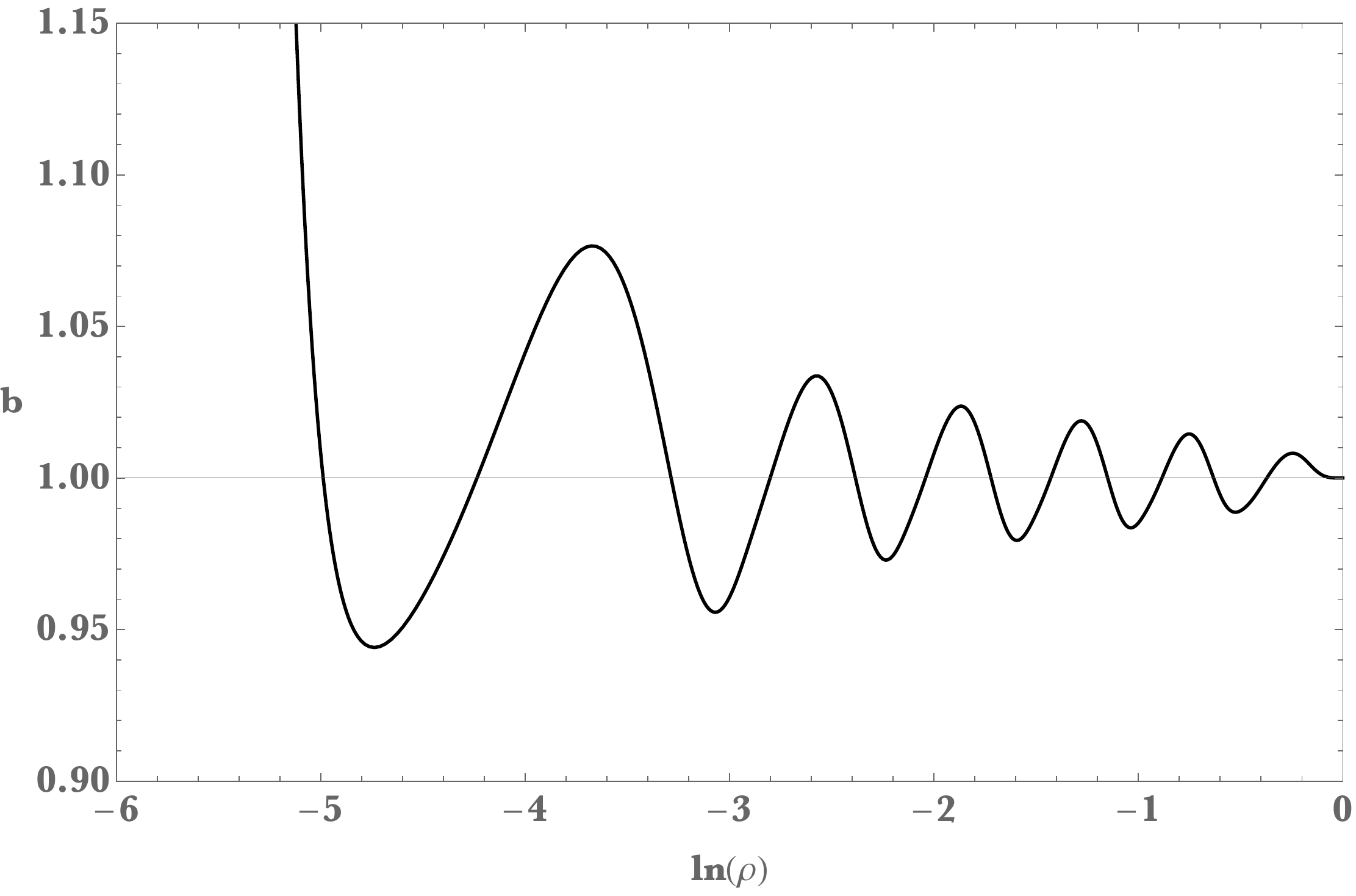}
\caption{Supercritical metric function $b$ in the bulk at horizon function.  Note the log scale showing self-similarity.  The portion of this configuration for $\ln(\rho)\lesssim-5$ becomes trapped behind a horizon, while the rest of the self-similar waveform propagates towards the boundary.}\label{fig:b}
\end{figure}

{\bf~Results and discussion --} In Fig.~\ref{fig:b}, we show the metric function $b$ at the time of apparent horizon formation for finely-tuned supercritical initial data.  We see that $b$ exhibits nearly equally spaced oscillations in $\ln \rho$. The self-similar structure in $b$ subsequently propagates outwards towards the boundary, where via (\ref{stress2}), it influences the evolution of the pressure anisotropy $\langle B \rangle$.

\begin{figure}
\centering
\includegraphics[width=.45\textwidth]{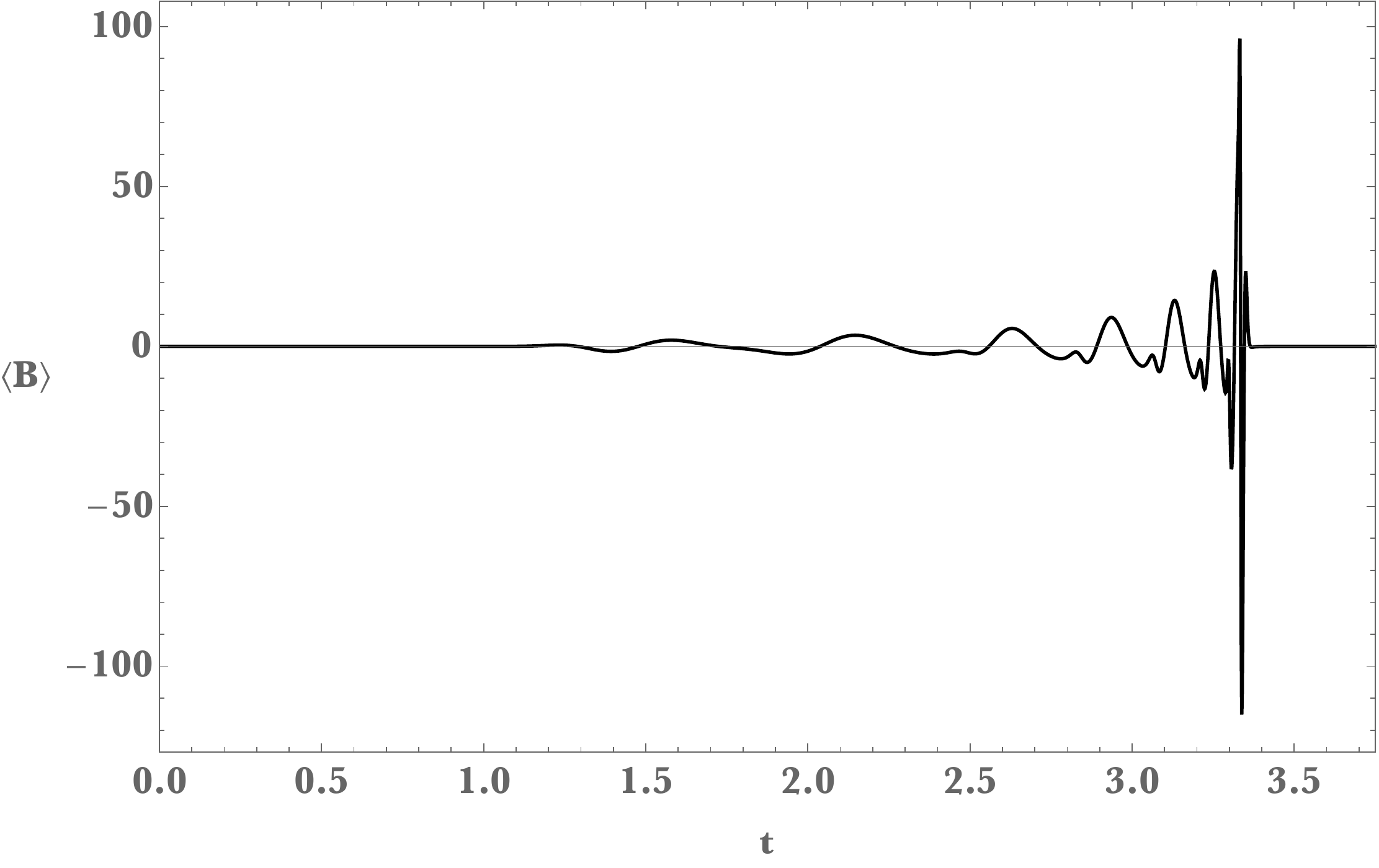}
\includegraphics[width=.45\textwidth]{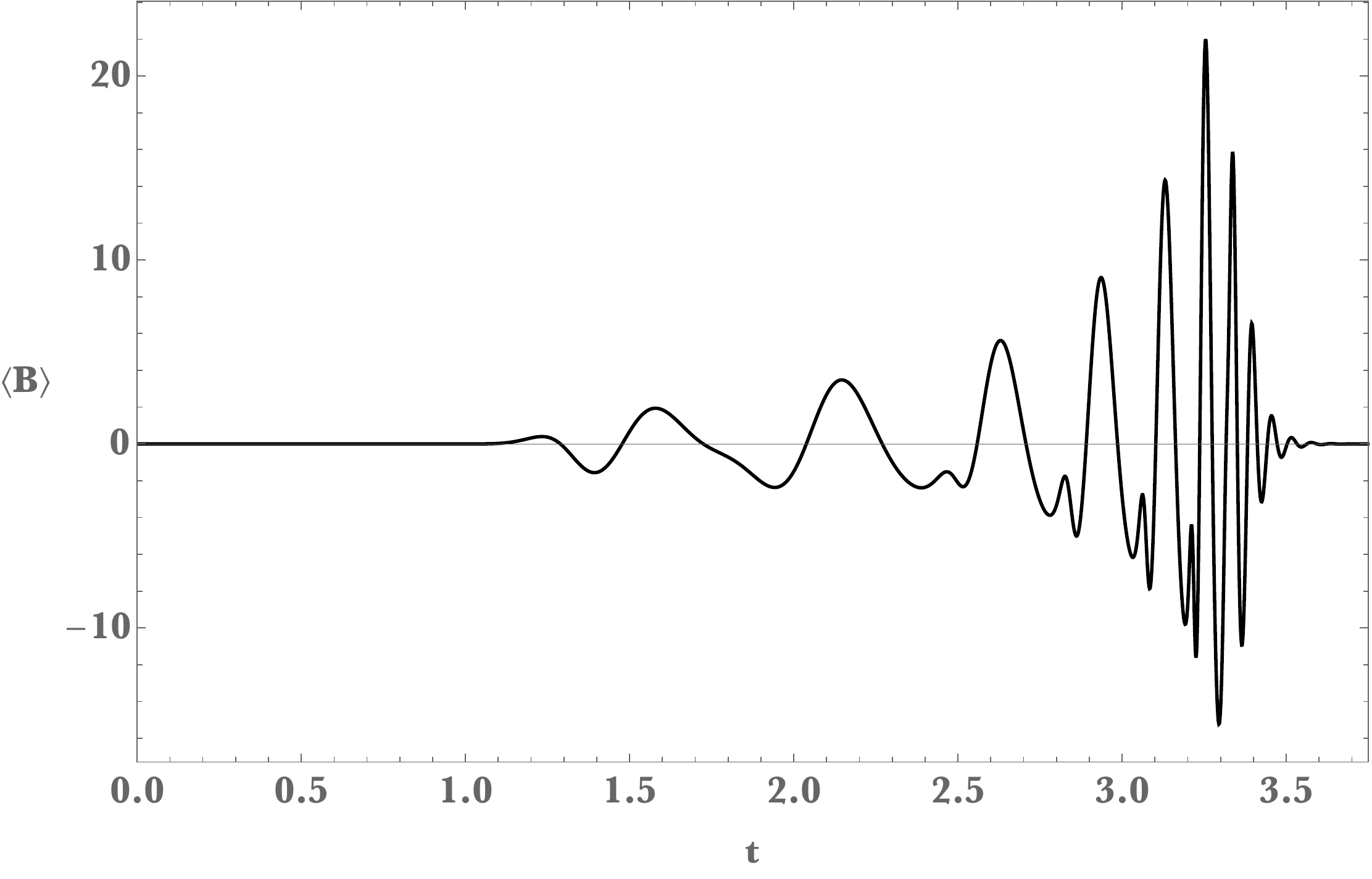}
\caption{Pressure anisotropy $\langle B\rangle$ as a function of boundary time $t$ for subcritical (top) and supercritical (bottom) initial data.  Initial data is fined tuned to criticality to about $1-\epsilon/\epsilon_*\sim 10^{-7}$.  For both data sets $\langle B \rangle $ oscillates with increasing frequency and amplitude as time progresses. For supercritical data the oscillations in $\langle B \rangle $ reach a maximum frequency $f_\mathrm{max}$, after which $\langle B \rangle $ attenuates in amplitude.
}\label{fig:opfcn}
\end{figure}

In Fig.~\ref{fig:opfcn}, we show $\langle B\rangle$ as a function of $t$ for one instance of subcritical and supercritical data. As the self-similar waveform in $b$ approaches the boundary, $\langle B\rangle$ begins to oscillate, with increasing frequency and amplitude as time progresses.  For subcritical data, the oscillations abruptly terminate at some maximum frequency $f_{\rm max}$.  In contrast, for supercritical data, the oscillations reach a maximum frequency $f_{\rm max}$ and then decay in amplitude while continuing to oscillate at frequency $\sim f_{\rm max}$. This is due to the newly formed horizon making causal contact with the boundary, with the subsequent attenuation of $\langle B \rangle$ reflecting the ringdown of the black hole.

\begin{figure}
\centering
\includegraphics[width=.45\textwidth]{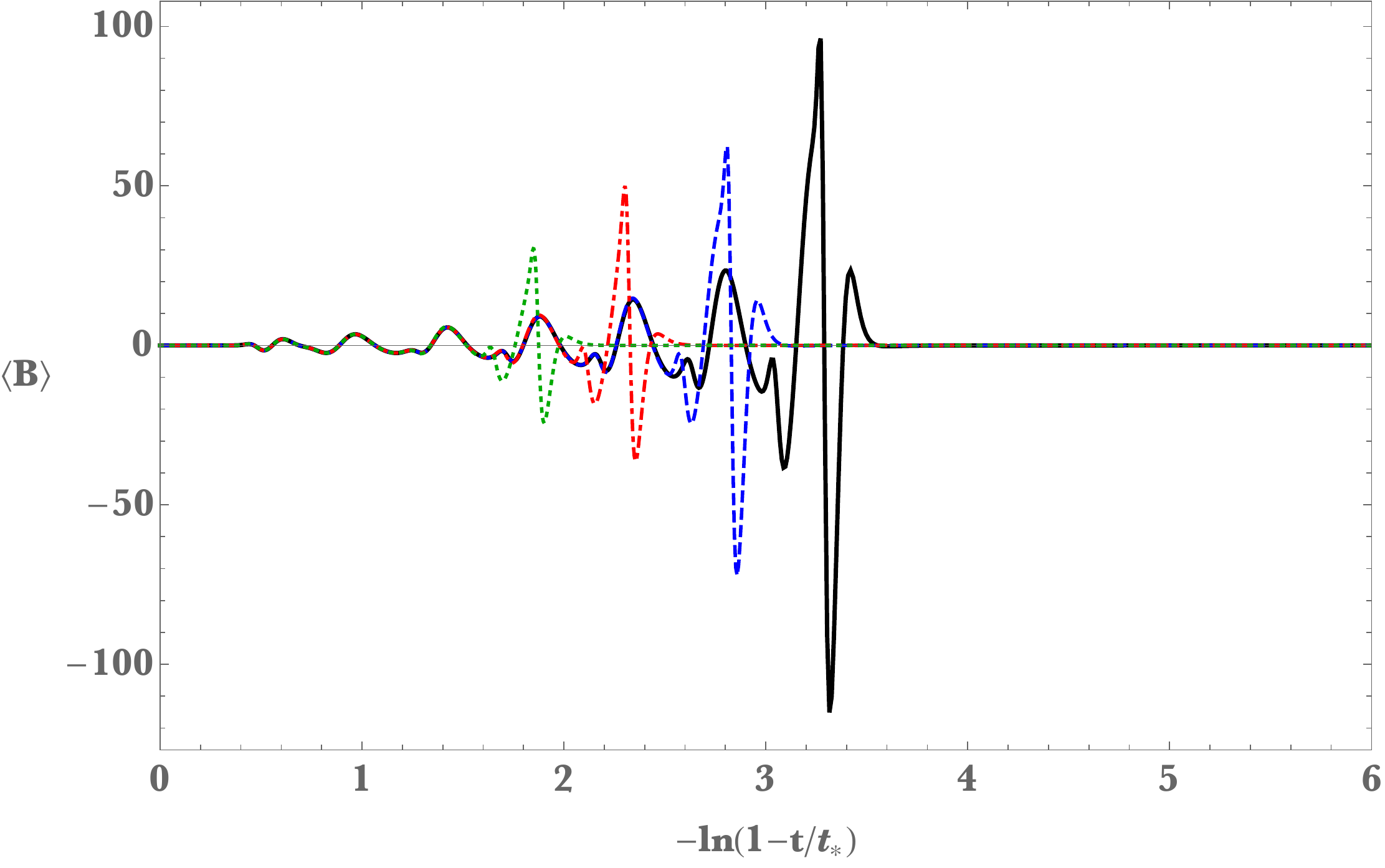}
\includegraphics[width=.45\textwidth]{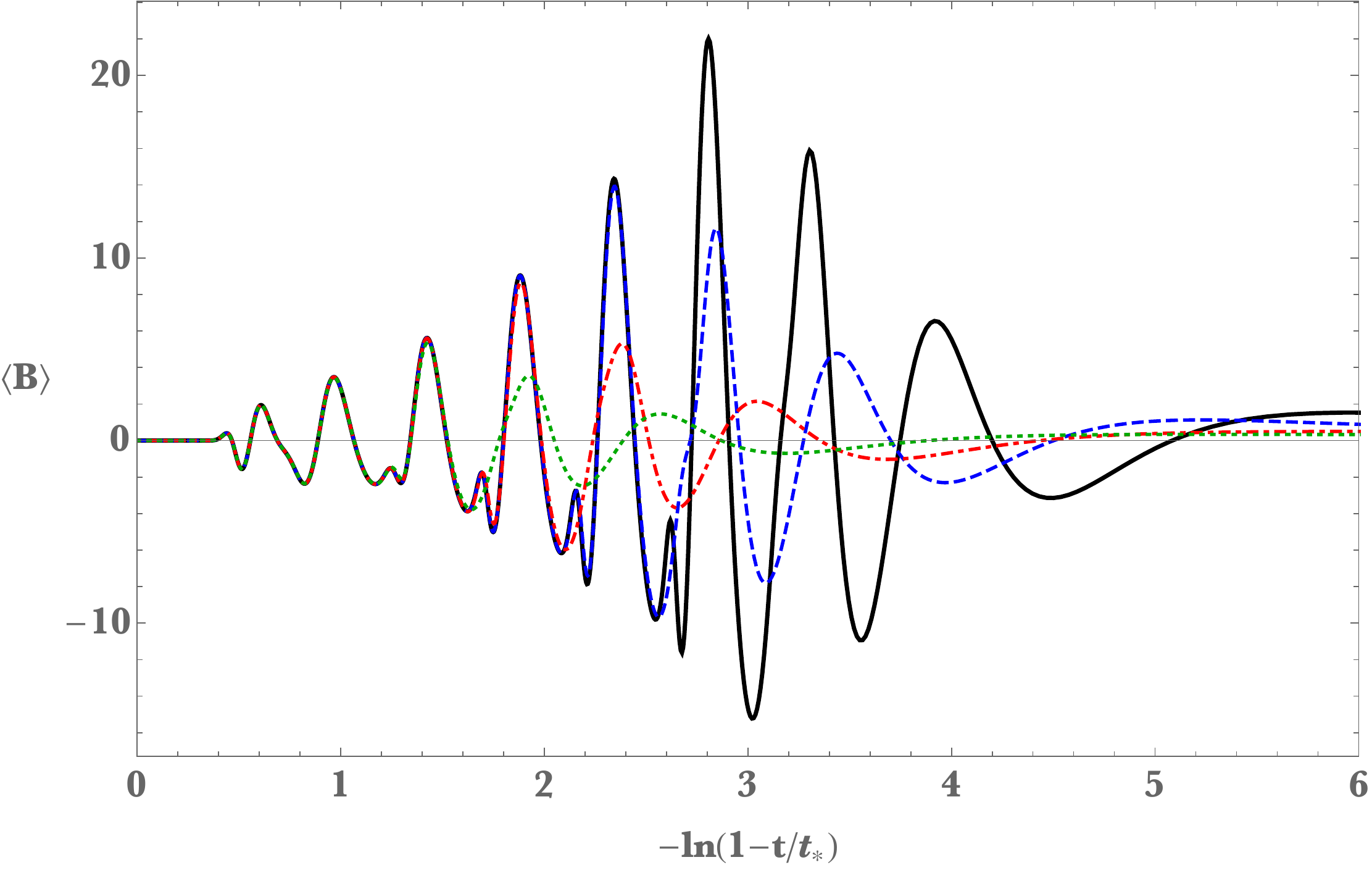}
\caption{Pressure anisotropy $\langle B\rangle$ as a function of $-\ln(1-t/t_*)$, where $t_*$ is chosen to preserve self-similarity, for subcritical (top) and supercritical (bottom) initial data.  Colored and dotted/dashed lines are less fine-tuned data.  Note that all curves agree at early times, and that successive echoes appear with more fine tuning.}\label{fig:opfcnlog}
\end{figure}

Both subcritical and supercritical simulations show a time $t_*$ about which $\langle B\rangle$ has self similar structure.  In Fig.~\ref{fig:opfcnlog}, we plot $\langle B\rangle$ as a function of $-\ln(1-t/t_*)$ for various amplitudes $\epsilon$.  We choose $t_*$ such that the oscillations in Fig. \ref{fig:opfcnlog} have constant period in $-\ln(t_*-t)$.  Note that at early times all curves in both plots overlap, and that fine-tuning closer to criticality adds more echos with successively increasing amplitudes.  The echoing period is $\Delta\approx0.46$, which is comparable to the flat space result \cite{Bizon:2005cp} of $\Delta\approx0.47$.  As extracting the echoing period involves a loss of fidelity due to fitting, our result for $\Delta$ is consistent with that of \cite{Bizon:2005cp}. Evidently, the self-similar structure in the gravity description gets imprinted on the boundary stress tensor $\langle T^{\mu}_{\ \nu}\rangle$. $t_*$ has the natural interpretation as the time at which the naked singularity of the critical solution makes causal contact with the AdS boundary.

The maximum frequency $f_{\rm max}$ reached by $\langle B\rangle$ at the termination of scale echoing can be related to the critical exponent $\gamma$.  For supercritical initial data, scale echoing terminates when a horizon is formed, where $f_{\rm max}$ has the same scale as the horizon radius: $f_{\rm max}\sim r_H^{-1}$.  Subsequently, the black hole should ring down with a frequency also set by this scale. But in five dimensions, $r_H \propto \sqrt{M_H}$, so the scaling relation (\ref{eq:scaling}) implies that
\begin{equation}
\label{eq:fmax}
f_{\rm max} \sim |\epsilon - \epsilon_*|^{-\gamma/2}.
\end{equation}

\begin{figure}
\centering
\includegraphics[width=.47\textwidth]{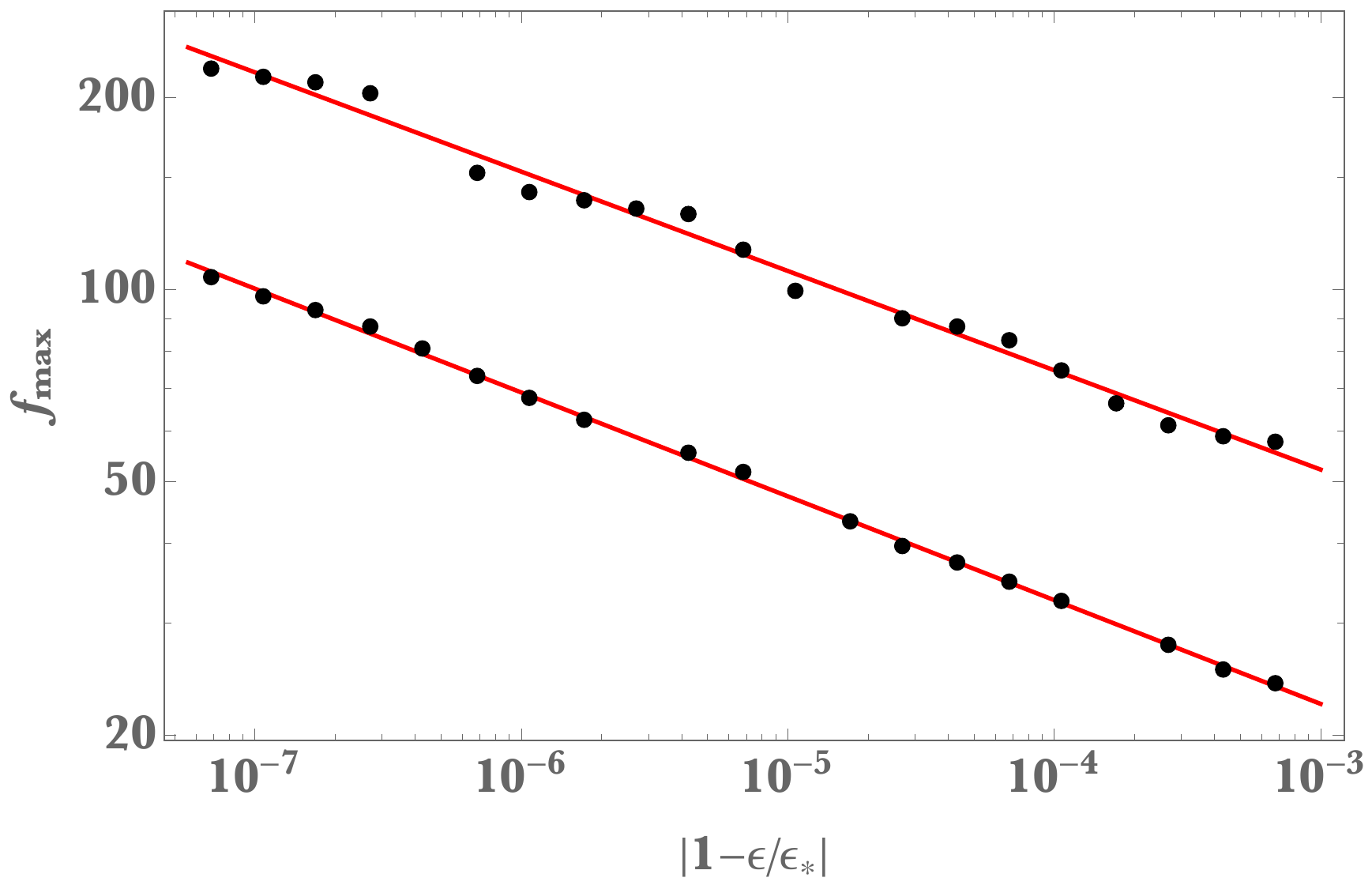}
\caption{Maximum echo frequency $f_{\rm max}$ of $\langle B \rangle$ for subcritical solutions (top) and supercritical solutions (bottom). The maximum echo frequency is reached when scale echoing in $\langle B \rangle$ terminates. Both subcritical and supercritical solutions scale as $f_{\rm max} \sim |\epsilon - \epsilon_*|^{-\gamma/2}$.}\label{fig:maxf}
\end{figure}

To test (\ref{eq:fmax}) on the boundary, we perform a short-time Fourier transform of $\langle B \rangle$ with Gaussian windowing.  We then locate the dominant peak in the windowed Fourier transform $f(t)$, which gives the local echo frequency of $\langle B \rangle$.  $f(t)$ takes its maximum value $f_{\rm max}$ when scale echoing terminates.  In Fig.~\ref{fig:maxf} we plot $f_{\rm max}$ as a function of $\epsilon$ for both subcritical (top) and supercritical (bottom) solutions \footnote{We have removed data points corresponding to runs with overly large constraint violation.}. The red lines in each plot are fits to (\ref{eq:fmax}).  For subcritical and supercrtical data the fits yield $\gamma = 0.32$ and $\gamma = 0.33$, respectively
\footnote{Due to the black hole ring down, there are more oscillations in $\langle B \rangle$ at $f \sim f_{\rm max}$ for supercritical solutions, allowing $f_{\rm max}$ to be determined with greater fidelity.}.
This should be compared to the flat space result of $\gamma = 0.33$ reported in \cite{Bizon:2005cp}.  Evidently, near the threshold of thermalization, the stress tensor oscillates at $f\sim f_{\rm max}$.


Note that the scaling (\ref{eq:fmax}) applies for both subcritical and supercritical data.  This result should be expected, since the critical exponent $\gamma$ is related to the only growing mode of perturbations of the critical solution, which is insensitive to whether the perturbation is supercritical or subcritical \cite{Gundlach:1996eg}.  Subcritical scaling has also be found in the maximum curvature at the origin \cite{Garfinkle:1998va}.

We also note that studies of critical collapse in asymptotically flat spacetime indicate the Bondi news function inherits the self-similar structure of the critical solution \cite{Purrer:2004nq}.  Presumably, scale echoing in the news function also terminates at maximum frequency $f_{\rm max}$.  In four dimensions, the horizon radius $r_H\sim M_H$, which means that scale echoing should terminate at $f_{\rm max} \sim |\epsilon - \epsilon_*|^{-\gamma}$.

\begin{figure}[t]
\centering
\includegraphics[width=.45\textwidth]{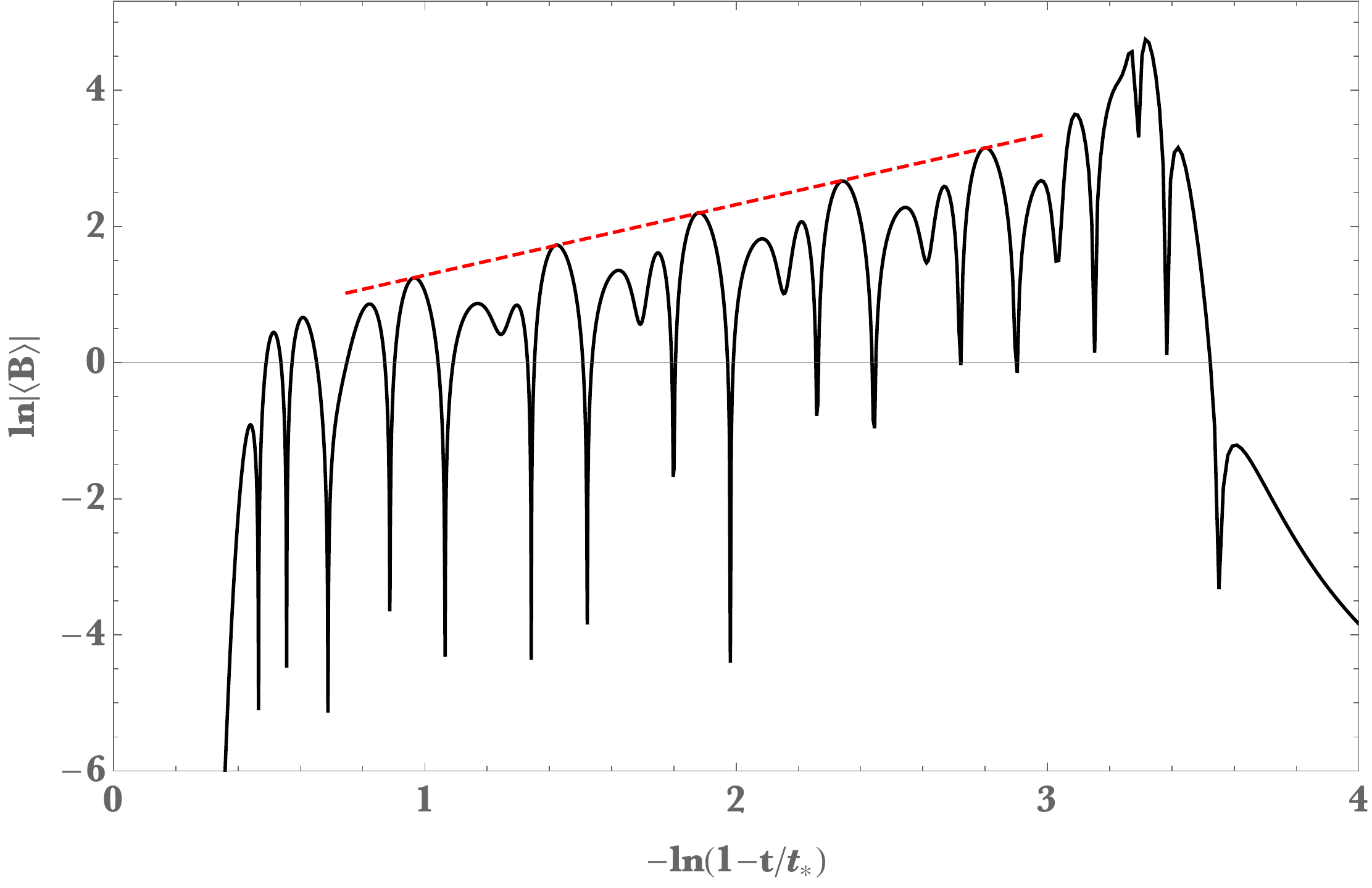}
\caption{$\ln|\langle B\rangle|$ as a function of $-\ln(1-t/t_*)$.  The dashed red line shows $1/(1-t/t_*)$.  Evidently, for the critical solution the pressure aniosotropy diverges when the naked singularity in the bulk makes causal contact with the boundary.}\label{fig:opfcnloglog}
\end{figure}

We now turn to the envelope of $\langle B \rangle$. In Fig.~\ref{fig:opfcnloglog}, we plot $\ln|\langle B\rangle|$ as a function of $-\ln(1-t/t_*)$. We see that the amplitude of each echo in $\ln|\langle B\rangle|$ appears linear in $-\ln(1-t/t_*)$ in the self-similar region.  (There are also transition regions to early times and times very close to $t_*$ which do not show self-similar behavior.) Our numerics are consistent with
\begin{equation}\label{powerlaw}
\langle B\rangle\propto 1/(1-t/t_*)\;,
\end{equation}
implying that the stress tensor diverges when the naked singularity makes causal contact with the boundary.

To understand how the envelope (\ref{powerlaw}) arises, we consider a linear analysis for the propagation of a self-similar waveform in $b$ to the boundary.  At linear order about AdS, the Einstein equation implies
\begin{equation}
\label{eq:linearanalysis}
b(t,\rho)=1+{\mathfrak{Re}}\sum_na_n e^{-i\omega_n t}\Psi_n(\rho),
\end{equation}
where $\omega_n \equiv 6+2n$ are the normal mode frequencies of AdS, $a_n$ are mode amplitudes, and
\begin{equation}
\Psi_n(\rho) \equiv \frac{1}{n}\rho^2(2{-}\rho^2)(1{-}\rho^2)^4 \, P^{(3,2)}_n(1 {-} 4 \rho^2 {+} 2 \rho^4)
\end{equation}
with $P^{(\alpha,\beta)}_n(x)$  Jacobi polynomials.
At large $n$ with $z \equiv n\rho$ fixed, $\Psi_n(\rho)\to\frac{1}{z \sqrt{2}} J_3(2 \sqrt{2} z)$ with $J_3$ a Bessel function.  One can then show that self-similarity at small $\rho$ requires that $n a_n$ be periodic in $\ln n$ at large $n$.

An expansion at the boundary gives $\Psi_n(\rho) = 8 n (-1)^n (1-\rho)^4$ for large $n$. Eq.~(\ref{stress2}) then implies the high frequency components of $ \langle B(t) \rangle$ read 
\begin{equation}
\label{eq:Bhifreq}
\!\!\!\!\! \textstyle \langle B(t) \rangle = \frac{\partial \mathcal A}{\partial t}, \  {\rm with} \
\mathcal A(t) \equiv \textstyle \frac{1}{4} {\mathfrak Im} \sum_{n} (-1)^n a_n e^{-i \omega_n t}.
\end{equation}
The near-origin self-similarity that requires periodicity of $n a_n$ in $\ln n$ therefore implies that $\mathcal A(t)$ is also self-similar (i.e. periodic in $-\ln (1-t/t_*)$ for some $t_*$).  By the chain rule, the envelope of $\langle B(t) \rangle$ is given by (\ref{powerlaw}). The $1/(1-t/t_*)$ envelope is therefore naturally interpreted as a consequence of self-similarity of the bulk geometry.


The divergence at $t_*$ assumes the validity of Einstein gravity, which will inevitably break down as the critical point is approached. In the limit $N \to \infty$ with $\lambda \gg 1$, higher derivative corrections to the Einstein-Hilbert action become important when the curvature scale and string scale $\ell_s$  become comparable.  In terms of CFT observables, this happens when  $f_{\rm max} \sim \lambda^{1/4}/L,$ where $L \ (=1 $ in our units) is the radius of the $S^3$ on the boundary.  This presumably means the pressure anisotropy ceases growing at time $t_*- t \sim L \lambda^{-1/4}$, meaning the maximum pressure anisotropy should scale like $\lambda^{1/4}$.  This suggests the critical phenomenon observed in this Letter doesn't exist at weak coupling. We also emphasize that the range of $\epsilon$ where critical phenomena can be observed can be made parametrically large by increasing $\lambda$.

It is striking that while we have started in a state that appears isotropic and close to equilibrium, the pressure anisotropy subsequently grows like $1/(1 - t/t_*)$, with self-similar oscillations, and then either abruptly vanishes (for subcritical initial conditions) or exponentially rings down (for super-critical initial conditions). To the best of our knowledge, this phenomenon is  entirely new in QFT. The ubiquity and universality of critical collapse suggests that similar behavior should be found in a wide variety of strongly coupled holographic QFTs.

{\bf~Acknowledgments --} It is a pleasure to thank Gary Horowitz, Laurence Yaffe, and our referees for helpful comments.   PMC is supported by the Black Hole Initiative at Harvard University, which is funded by a grant from the John Templeton Foundation.  BW is supported by NSERC.  We thank the Yukawa Institute for Theoretical Physics at Kyoto University, where this work was initiated during the YITP-T-18-05 workshop on ``Dynamics in Strong Gravity Universe."

\bibliography{refs}
\bibliographystyle{utphys}
\end{document}